\documentclass{article}
\usepackage{epsfig}
\title{Observation of $K_S^0$ semileptonic decays with CMD-2 detector.
\thanks{This work is supported in part by the Russian
Foundation for Basic Research under grant RFBR-98-02-17851 
and the US DOE grant DEFG0291ER40646.}}

\author{
R.R.Akhmetshin,  E.V.Anashkin,  M.Arpagaus,  V.M.Aulchenko,  \and
V.Sh.Banzarov,  L.M.Barkov,  S.E.Baru,  N.S.Bashtovoy,  \and
A.E.Bondar,  D.V.Bondarev,  A.V.Bragin,  D.V.Chernyak,  \and
A.G.Chertovskikh,  A.S.Dvoretsky,  S.I.Eidelman,  G.V.Fedotovich,  \and
N.I.Gabyshev,  A.A.Grebeniuk,  D.N.Grigoriev,  P.M.Ivanov,  \and
S.V.Karpov,  B.I.Khazin,  I.A.Koop,  L.M.Kurdadze,  A.S.Kuzmin,  \and
I.B.Logashenko,  P.A.Lukin,  K.Yu.Mikhailov,  I.N.Nesterenko,  \and
V.S.Okhapkin,  E.A.Perevedentsev,  A.S.Popov,  T.A.Purlatz,  \and
N.I.Root,  A.A.Ruban,  N.M.Ryskulov,  A.G.Shamov,  \and
Yu.M.Shatunov,  A.I.Shekhtman,  B.A.Shwartz,  V.A.Sidorov,  \and
A.N.Skrinsky,  V.P.Smakhtin,  I.G.Snopkov,  E.P.Solodov,  \and
P.Yu.Stepanov,  A.I.Sukhanov,  V.M.Titov, Yu.V.Yudin,  S.G.Zverev \\ \vspace{2mm}
{\it Budker Institute of Nuclear Physics, Novosibirsk, 630090,
Russia} \\ \vspace{2mm}
J.A.Thompson \\ \vspace{2mm}
{\it University of Pittsburgh, Pittsburgh, PA 15260, USA}
}

\date{}

\begin{document}
\maketitle
\begin{abstract}
The decay $K_S^0 \to  \pi e \nu$ has been observed by the CMD-2
detector at the $e^{+}e^{-}$ collider VEPP-2M at Novosibirsk. Of 6 
million produced $K_L^0K_S^0$ pairs, $75 \pm 13$ events of the 
$K_S^0 \to \pi e \nu$ decay were selected. The corresponding branching 
ratio is B($K_S^0 \to \pi e \nu$)=$(7.2 \pm 1.4)\times10^{-4}$. This 
result is consistent with the evaluation of B($K_S^0 \to \pi e \nu$) 
from the $K_L^0$ semileptonic rate and $K_S^0$ lifetime assuming 
$\Delta S=\Delta Q$ . 
\end{abstract}
\section{Introduction}
While semileptonic decays of the K$_L^0$ have been well measured, 
the information on similar decays of K$_S^0$ is scarce. The
only measurement of the K$_S \to \pi^{\pm} e^{\mp} \nu$ performed 
long ago assumed $\Delta S=\Delta Q$ and has low accuracy \cite{aubert}. 
The Review of Particle Physics evaluates the corresponding decay rate 
indirectly, using the K$_L^0$ measurements and assuming 
that $\Delta$S=$\Delta$Q so that \\
$\Gamma(K_S^0 \to \pi^{\pm} e^{\mp} \nu) = 
~\Gamma(K_L^0 \to \pi^{\pm} e^{\mp} \nu)$ \cite{pdg}. \\
   We present results of the direct measurement of the branching ratio for 
the $K_S^0 \to \pi e \nu$ using the unique opportunity to study events
containing a pure $K_L^0K_S^0$ state produced in the reaction 
$e^{+}e^{-} \to \phi \to K_L^0K_S^0$. The data were collected during the 
period of 1993-1998 with the CMD-2 detector \cite{cmd2,cmd3}.

The CMD-2 is a general purpose detector consisting of a drift chamber (DC) 
and proportional Z-chamber (ZC) used for the trigger, both inside a thin 
$(0.4 X_0)$ superconducting solenoid with a field of 1 T. Outside the field, 
there is a barrel (CsI) calorimeter and a muon range system. 
The CsI calorimeter covers polar angles from 0.8 to 2.3 radian. 
The vacuum beam pipe with a radius of 1.8 cm is placed inside the DC and 
$K_S^0$ mesons with the decay length $\lambda =0.6$ cm decay within it. 
The DC has momentum resolution of 3\% for 200 MeV/c charged particles.
The CsI calorimeter with $6\times6\times15$ $ cm^3$ crystals is placed at 
a distance of 40 cm from the beam axis and about a half of $K_L^0$ mesons with 
the decay length $\lambda =3.3$ m 
has interactions within CsI crystals. The energy resolution for photons in the 
CsI calorimeter is about 8\%. Charged particles from the neutral kaon decays 
have momenta less than 290 MeV/c and stop within the  CsI crystals. 
\section{Analysis}
$K_S^0$ decays can be tagged using the presence of the second vertex
with two charged particles at a distance from the $e^+e^-$ 
interaction region or the CsI cluster from $K_L^0$ interactions in CsI. 
The most probable decay channel $K_S^0 \to \pi ^+ \pi ^-$ was used for the
normalization of the semileptonic $K_S^0 \to \pi e \nu$ decay. Both channels 
have a vertex with two charged particles near the beam axis. 

\begin{figure}[tbh]
  \begin {minipage}[t] {0.48\linewidth}
  \centering\includegraphics [width=1.0\linewidth] {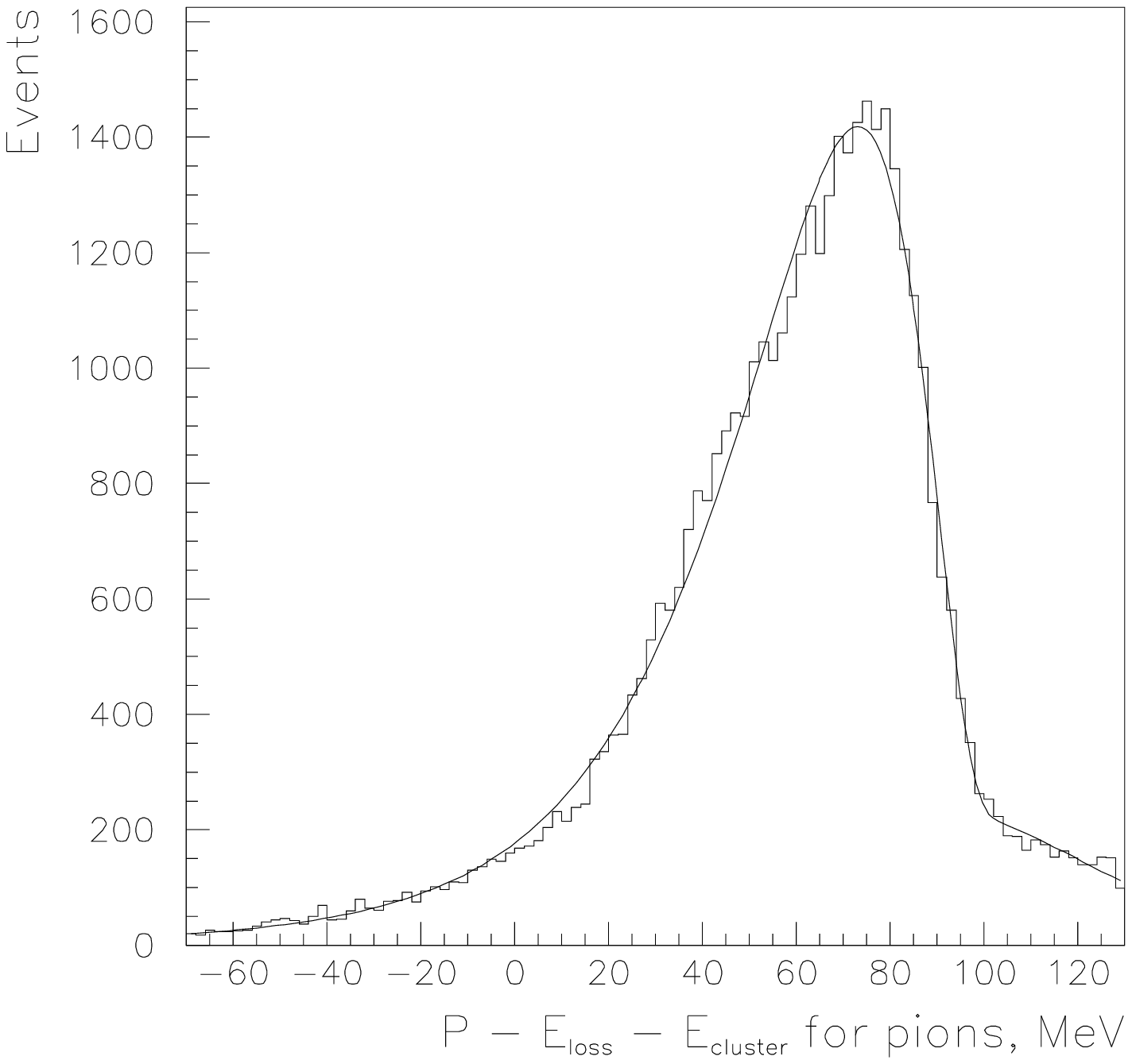}
  \caption {DPE distribution for pions with momenta 160--200 MeV/c in 
  $K_S^0 \to \pi^+ \pi^-$ decay.} 
  \label{picrys3}
  \end{minipage}
  \hspace{0.02\linewidth}
  \begin{minipage}[t] {0.48\linewidth}
  \centering\includegraphics [width=1.0\linewidth] {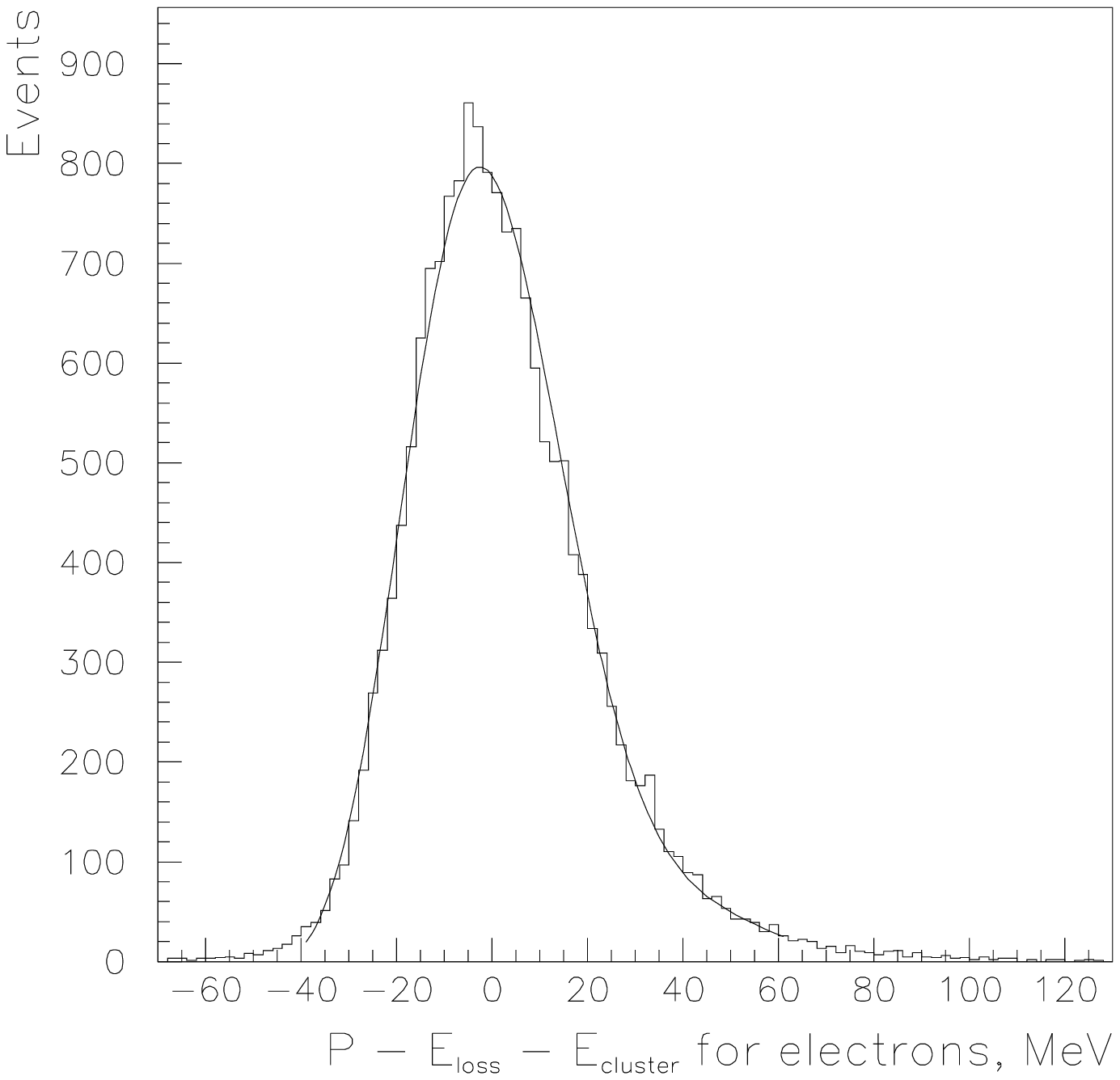}
  \caption {DPE distribution for collinear electrons with momentum 
   195 MeV/c.}
  \label{picrys1}
  \end{minipage}
\end {figure}

\begin{figure}[tbh]
  \begin {minipage}[t] {0.48\linewidth}
  \centering\includegraphics [width=1.0\linewidth] {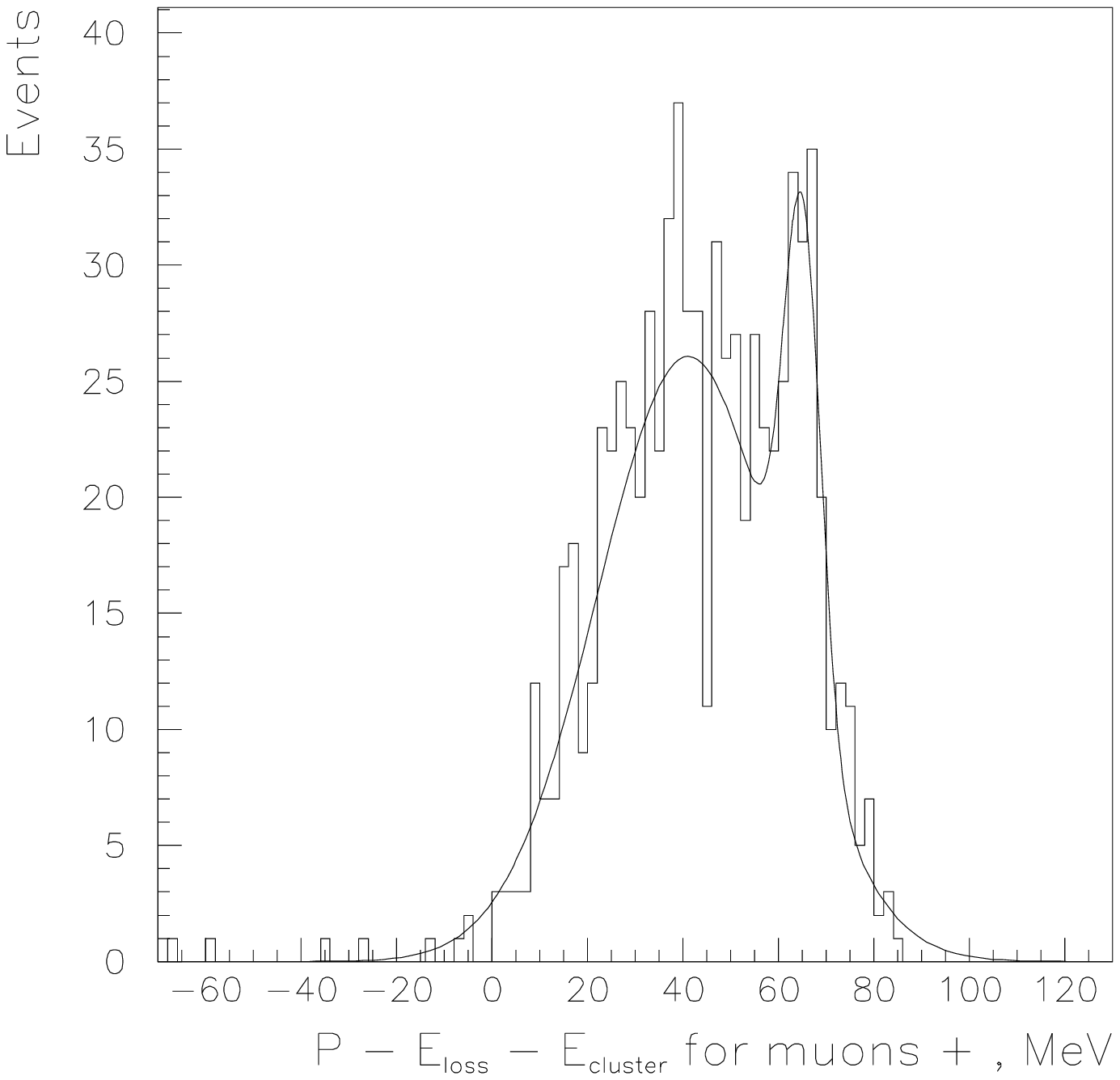}
  \caption {DPE distribution for positive muons with momentum 164 MeV/c.}
  \label{picrys2+}
  \end{minipage}
  \hspace{0.02\linewidth}
  \begin{minipage}[t] {0.48\linewidth}
  \centering\includegraphics [width=1.0\linewidth] {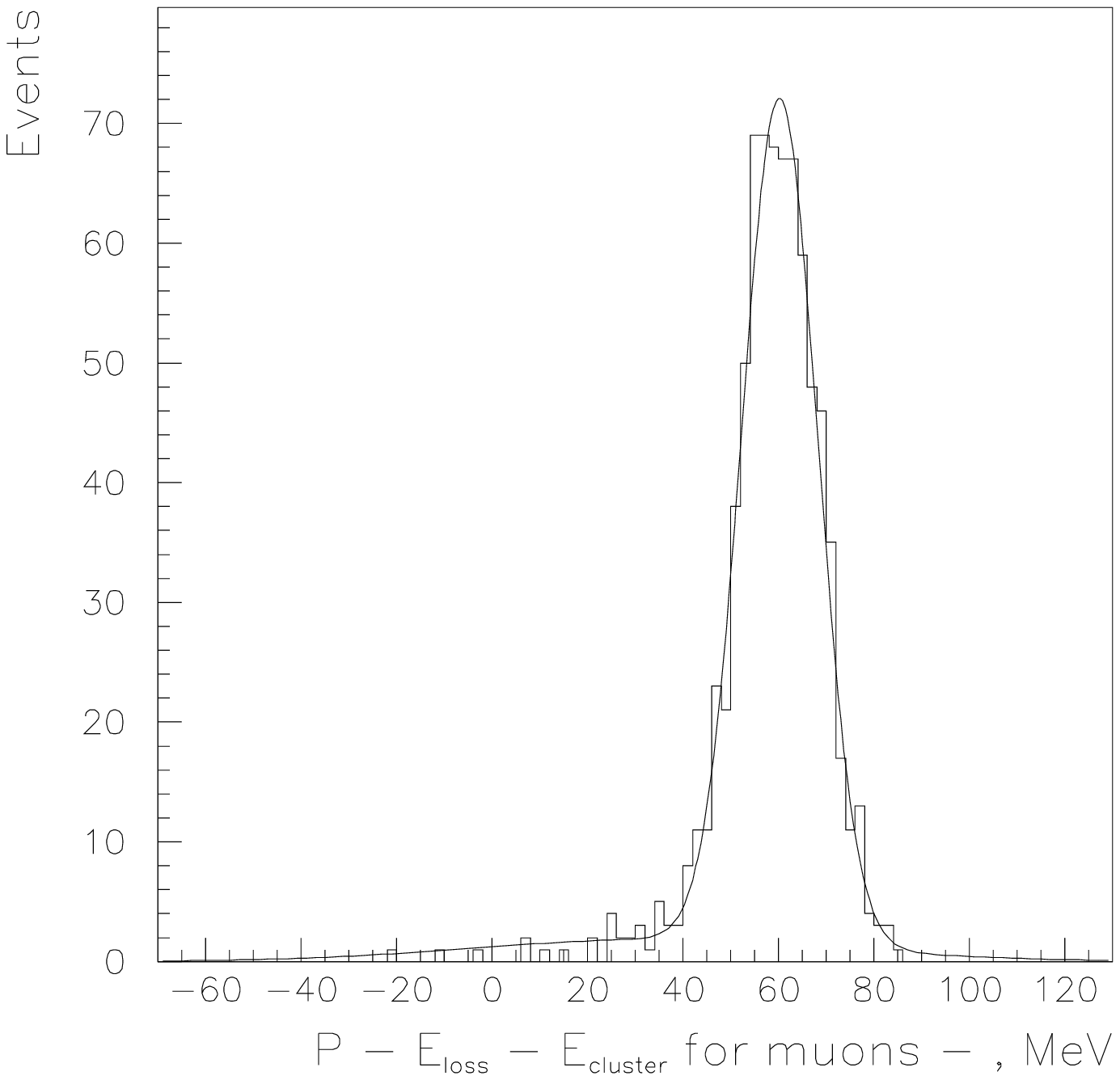}
  \caption {DPE distribution for negative muons with momentum 164 MeV/c.}
  \label{picrys2-}
  \end{minipage}
\end {figure}

To identify electrons in the decay under study, we are using the 
difference between measured momentum and energy loss in 
the detector material for stopped particles.
The basic parameter used for charged particle  identification was
\[ DPE =P_{particle} - E_{loss} - E_{cluster}, \]
where $P_{particle}$ is the particle momentum measured in the DC, 
$E_{loss} $ is the average ionization energy loss (about 10 MeV) in the 
material in front of the CsI calorimeter, $E_{cluster}$ is the energy 
deposition in the CsI cluster matched with a particle track.  CsI 
clusters which do not match any track are further referred to as photons. 
Electrons must have DPE=0 if the resolution of the detector is ideal and the
leakage of showers in the CsI calorimeter is negligible. On the other hand, 
pions and positive muons have a broad distribution
displaced from zero. Negative muons have a sharp peak 
displaced from zero as the energy of the CsI cluster is equal to the
difference between the muon kinetic energy and $E_{loss}$. Pions 
from the decay $K_S^0 \to \pi ^+ \pi^-$ were used to obtain
the distribution over this parameter for charged  pions in the
momentum range 160 - 200 MeV/c. This distribution together with the fitting 
function is shown in Fig.\ref{picrys3}. For electrons and
muons the distribution over this parameter was obtained from experimental 
data for reactions $e^+e^- \to e^+e^-, \mu ^+\mu ^-$ at the beam energy 
of 195 MeV. At this energy particle momenta are 195 MeV/c for electrons and 
164 MeV/c for muons. The DPE distribution for electrons as well as the 
fitting function are shown in Fig.\ref{picrys1}. The same distribution for 
muons  (Fig.\ref{picrys2+} and Fig.\ref{picrys2-}) overlaps with 
the distribution for pions and this is properly taken into account. 

\hspace*{6 mm}Some kinematic features for the decay mode $K_S^0 \to \pi e \nu$ are : 

\begin{itemize}
\item The opening angle between two tracks is between 0 and $\pi$ 

\item The  momenta of charged particles are less than 290 MeV/c 

\item The total energy of charged particles (assuming that both particles are 
      charged pions) is between 330 and 550 MeV. 

\hspace*{6 mm}The same parameters for the decay mode $K_S^0 \to \pi^+ \pi^-$ are : 

\item The opening angle between two tracks is more than 2.6 radians 

\item The pion momenta are between 160 and 270 MeV/c 

\item The total energy of charged particles (assuming that both particles are 
      charged pions) is equal to the beam energy (between 508 and 512 MeV). 

\hspace*{6 mm}The selection criteria for both modes of $K_S^0$ decay were:

\item One or two vertices are found in the event 

\item Two minimum ionizing tracks with the opposite charge sign are 
reconstructed from the first vertex (nearest to the beam) and there is 
no other track with distance to the beam less than 1.4 cm 

\item The distance from the first vertex to the beam is between 
0.2 cm and 1.4 cm. This cut rejects background from the beam region and 
material of the beam pipe 

\item The distance from the first vertex to the interaction point along 
the beam direction is less than 7 cm 

\item Each charged particle at the first vertex has a momentum between 
90 and 270 MeV/c since particles with a momentum less than 90 MeV/c can 
not  reach the CsI   calorimeter in the magnetic field of the detector 

\item Each track from the first vertex crosses all sensitive layers in the 
DC in the radial direction and therefore has a polar angle $\theta$ 
between 0.87 and 2.27 radians 

\item Each charged particle at the first vertex fires the ZC and does not 
fire the muon range system 

\item The azimuthal angle difference between two tracks at the first 
vertex ($\Delta \phi $) is between 0.17 and 2.97 radians 

\item The azimuthal angle difference ($\Delta \phi $) between 
the plane ``the first vertex ($K_S^0$) --- the beam axis'' and 
a photon  with the energy greater than 50 MeV (supposedly the $K_L^0$
cluster) or the second vertex in DC (supposedly the $K_L^0$ decay in DC)
is within $\pm 0.5$ radian 

\item There are no photons with the energy greater than 15 MeV 
outside the direction between ``the first vertex --- the beam axis'' 
$\pm 1$ radian in the $\phi $-plane. 
      This cut rejects background from processes with the neutral pions. 

\hspace*{6 mm}To select the decay mode $K_S^0 \rightarrow \pi e \nu$ the 
following criteria for the first vertex were additionally used: 
\item The opening angle between two tracks is between 0.35 and 2.50 radians 

\item The total energy of charged particles (assuming that both particles are 
      charged pions) is between 300 and 470 MeV 

\item The DPE parameter corresponding to the charged particle at the first 
vertex is included into a histogram when this particle track  
matches the CsI cluster independently of the matching conditions of the
other track 

\item The invariant mass squared of the assumed neutrino is greater than 
     $-10000$ $MeV^2 /c^4$ and less than 6000 $MeV^2 /c^4$. 

\hspace*{6 mm}To select the decay mode $K_S^0 \rightarrow \pi^+ \pi^-$ the following 
criteria for the first vertex were additionally used: 

\item The opening angle between two tracks is more than 2.55 radians 

\item The total energy of charged particles (assuming that both particles are 
      charged pions) is between 480 and 540 MeV 

\item The pion momenta are between 140 and 270 MeV/c 

\item The average momentum of two charged pions is between 190 and 230 MeV/c 

\item The ratio of the smaller momentum to the larger one is 
more than 0.58  

\item The angle between the vector sum of momenta and the direction 
``the beam axis $\rightarrow$ the first vertex'' is less than $\pi$/2 

\item Each charged particle at the first vertex has a matched CsI cluster 

\item The invariant mass squared of the assumed photon is greater 
than $-10000$ $MeV^2 /c^4$   and less than 6000 $MeV^2 /c^4$.
\end{itemize}

\section{Results and discussion}

\begin{figure}[tbh]
  \begin {minipage}[t] {0.48\linewidth}
  \centering\includegraphics [width=1.0\linewidth] {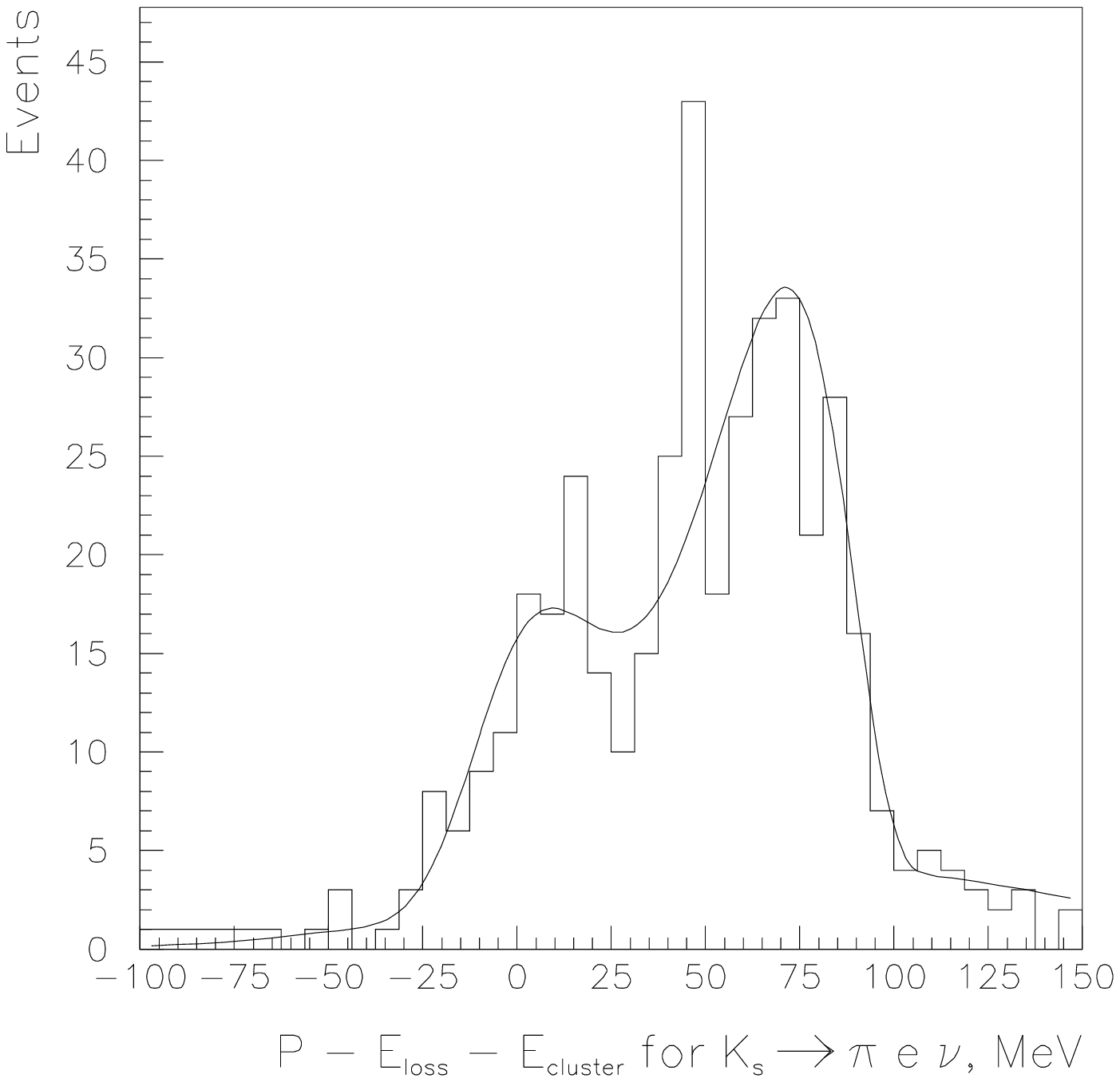}
  \caption {DPE distribution for charged particles in $K_S^0$ decays.}
  \label{picrys4}
  \end{minipage}
  \hspace{0.02\linewidth}
  \begin{minipage}[t] {0.48\linewidth}
  \centering\includegraphics [width=1.0\linewidth] {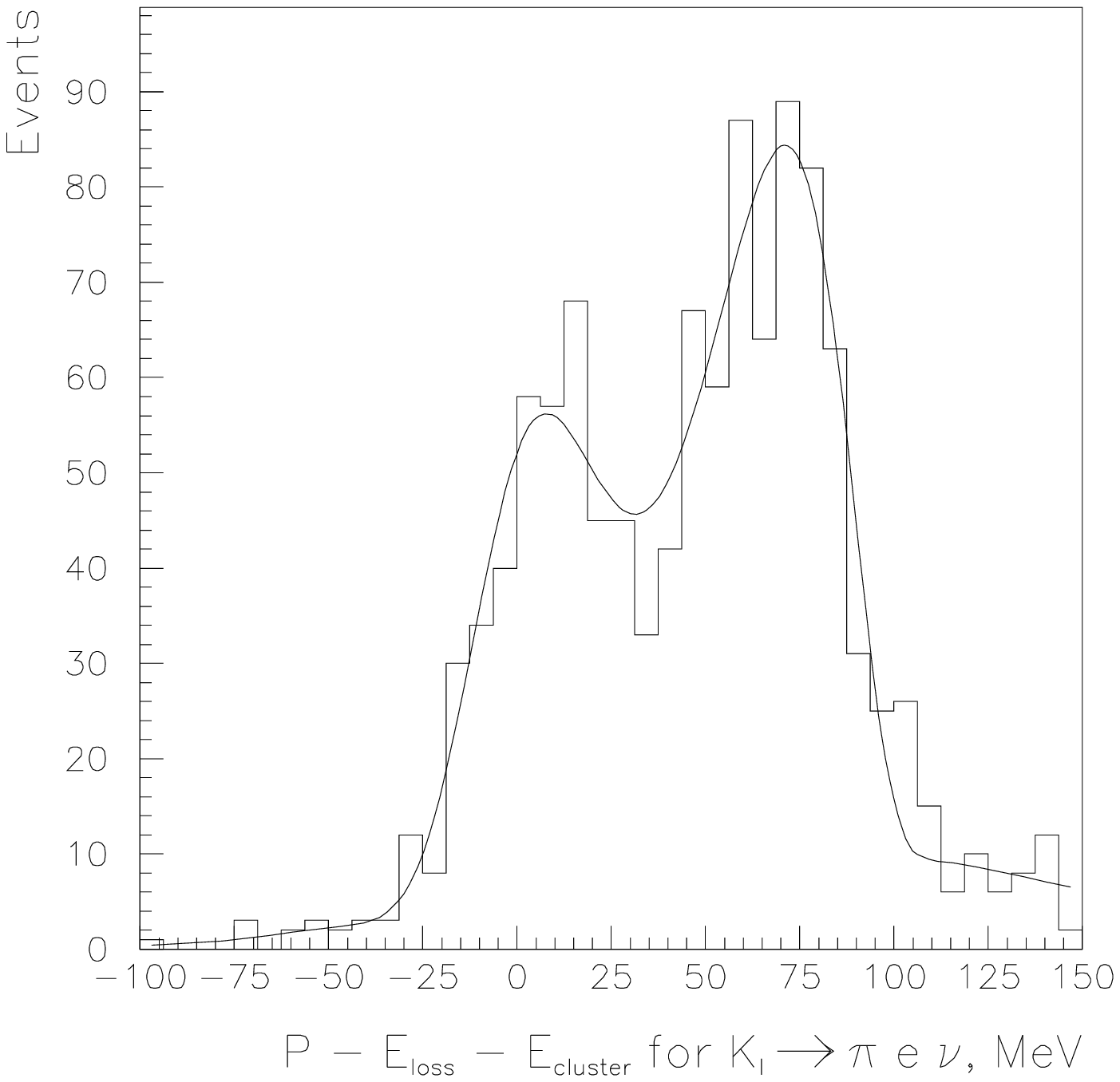}
  \caption {DPE distribution for charged particles in $K_L^0$ decays.}
  \label{picrys5}
  \end{minipage}
\end {figure}

The DPE distribution for events selected as candidates for the decay $K_S^0 \to \pi e \nu$ 
is shown in Fig.\ref{picrys4}. The data were fit using the DPE distribution of 
e, $\mu$ and $\pi$ measured in experiment. The result of the fit for the 
number of the electrons is $N_e = 83.5 \pm 12.7$. The number of mesons is 
equal to $N_m = 354 \pm 21$. The distribution over the 
distance between the vertex and beam axis for these events is consistent
with that for $K_S^0 \to \pi^+ \pi^-$ decays. 

The main background for the $K_S^0 \rightarrow \pi e \nu$ decay mode 
after applying the above cuts comes from the decays 
$K_S^0 \rightarrow \pi^+ \pi^- \gamma$, $K_S^0 \rightarrow \pi \mu \nu$ 
and $K_L^0\rightarrow \pi e \nu$. The former two processes are taken 
into account while fitting the histogram over DPE (the fit has two free
parameters - the number of electrons $N_e$ and that of mesons
$N_{\pi}$ + $N_{\mu}$).
To take into account the background from the latter process, 
the same procedure was applied to events with a distance from the 
first vertex to the beam axis between 3 and 7 cm. The resulting 
number of electrons for these events is $24.2 \pm 6.1$. Taking into 
account the dependence of the efficiency of vertex reconctruction 
on the distance from the beam axis as well as the ratio of the 
distance intervals for $K_L^0$- and $K_S^0$- decays it was found that 
the contribution of the $K_L^0$-decays is equal to $8.6 \pm 2.2$. 
Thus, the number of electrons and correspondingly the number of events
of the $K_S^0 \to \pi e \nu$ decay, is equal to
\[ N_e = 75 \pm 13 . \]

To illustrate the correctness of the identification based on the
DPE parameter, we applied the same procedure to look for events
of the decay $K_L^0 \to \pi e \nu$. Events were
selected in which  there were a $K_S^0 \to \pi^+ \pi^-$ 
decay near the beam axis and a second vertex in DC at a long distance 
from the beam axis. The ratio of the number of electrons $N_e$ to the 
number of pions and muons $N_{\mu} +N_{\pi}$ can be calculated from 
the branching ratios of the main decay modes of the $K_L^0$ and should 
be equal to 0.33 \cite{pdg}. The DPE distribution for events 
selected as candidates for the decays of $K_L^0$ meson is shown 
in Fig.\ref{picrys5}. The results of the fit are: $N_e = 300 \pm 22$, 
$N_m = 887 \pm 34$ and their ratio is 0.34 $\pm$ 0.03 in agreement with
the estimate above.  

Under the applied cuts the number of $K_S^0 \rightarrow \pi^+ \pi^-$ 
detected decays equals $N_{\pi\pi}=178110$ (of about 6 million 
produced $K_S^0K_L^0$ pairs). Applying the selection criteria above to the 
events from simulation, one obtains that the ratio of the detection 
efficiency for the normalization process to one for the process under study 
should be $\varepsilon _{rel}=2.49 \pm 0.12$.
The simulation of the $K_S^0 \to \pi e \nu$ decay was performed using the same 
Dalitz plot as for the $K_L^0$ decay. Then one would expect for the branching 
ratio 
\[B(K_S^0 \to \pi e \nu)=(N_e\cdot \varepsilon_{rel}/N_{\pi\pi})\cdot 
B(K_S^0 \to \pi^+ \pi^-).\]
From 75 $\pm$ 13 events observed by us
the following result was obtained  for the branching ratio: 

\[B(K_S^0 \to \pi e \nu)=(7.2 \pm 1.4)\times10^{-4}.\] 

The quoted error contains the statistical error 
and the systematic uncertainty (5\% from the simulation detection
efficiency and 5\% from the selection criteria) added in quadrature.
This result is consistent with the previous determination 
of B($K_S^0 \to \pi e \nu$) \cite{aubert} as well as with the
Review of Particle Physics evaluation \cite{pdg}.

\section{Conclusions}
The possibility to perform measurements using events containing the pure 
$K_L^0K_S^0$ state is available at the VEPP-2M collider in Novosibirsk. 
This circumstance permits to have tagged $K_S^0$ mesons in contrast to 
experiments with the neutral kaons, produced by the charge exchange 
of charged kaons at a target. Using the difference between measured 
momentum and energy loss in the detector material for stopped 
particles, the electrons from $K_S^0$ decays were identified. 
The branching ratio is B($K_S^0 \to \pi e \nu$)=$(7.2 \pm 1.4)\times10^{-4}$. 
It is the first direct measurement of this quantity.  

The value of the branching ratio does not contradict to that, calculated 
from $K_L^0$ semileptonic rates and $K_S^0$ lifetime assuming $\Delta S=\Delta Q$. 

{\bf Acknowledgements\/}

We would like to thank the technical personnel of VEPP-2M for the machine and 
detector support during experimental runs.

\end{document}